# Effect of molar volume on the interaction between mobile ions in mixed alkali oxyfluoro vanadate glasses: An Electrical modulus analysis.

Gajanan V Honnavar[1#*]

[1] Department of Physics, College of Science, Bahir Dar University , Bahir Dar,P.O. - 79 Ethiopia

# On sabbatical from PES Institute of Technology – Bangalore South Campus, Near Electronic City, Hosur Road, Bangalore – 560 100, India.

*gajanan.honnavar@gmail.com

Telephone: +91 080 66186610

Fax: +91 080 28521630

**Abstract**

Electrical modulus analysis of the frequency dependent conductivity data over the range from 100 Hz to 10 MHz and at temperature range of 140 $^{o}$C to 300 $^{o}$C were carried out on the mixed alkali glass samples having general formula $40V_2O_5 - 30BaF_2 - (30-x) LiF - xRbF$ where x = 5, 10,15,20,25 and 30. The analysis of the data is done using Kohlrausch-Williams-Watts (KWW) equation which describes the relaxation behavior of the mobile species. We have used the Bergman's equation (R. Bergman, J. Apl. Phys. **88** (2000) 1356.) for fitting the imaginary part of the electrical modulus. The stretching parameter $\beta$ derived from this fit shows a nonlinear variation with respect to alkali content which is attributed to mixed alkali effect. The variations of $\beta$ with respect to alkali element concentration is analysed on the basis of alkali ion distance and molar volume. It is found that the coupling between the mobile ions is better at 50 mol% of the alkali mixture when compared to single alkali concentrations. It is observed that coupling between the ions are weak in case of all Rubidium glass than all Lithium or Lithium-Rubidium mixture, even though the alkali concentration is on the higher (30 mol %) side. The contributions from fast and slow types of relaxation mechanisms were determined. It again shows that at low concentrations of Rubidium ions, there exists a good coupling between alkali ions and contribute to a macroscopic relaxation time responsible for the stretching, whereas at high concentrations of Rubidium ions the relaxing sites decouple and relax with a microscopic relaxation time. A scaling of modulus spectra with respect to frequency shows that stretching parameter $\beta$ is slightly dependent on temperature.

## 1. Introduction

Electrical transport in disordered systems like glasses and polymers is an important and active field of research [1,2]. Glasses were used as fast ionic conductors in the past [3]. Recently glasses have found to be useful as battery materials [4], and as transparent solar energy harvesters [5].

The ionic transport in glasses is usually studied using impedance spectroscopy [6]. The predominant phenomenon of relaxation, in the frequency range of few hundred Hz to few MHz, is through molecular orientation and reorientation; and hence these ranges of frequencies are normally used for the study. The AC conductivity spectra can be represented in different ways like complex admittance, complex permittivity, complex impedance and complex electrical modulus [1].

Electrical modulus representation of the ac conductivity data, originally introduced by McCrum et. al [7], but extensively applied to study vitreous ionic conductors by Moynihan and co-workers [8–10], has gained popularity by settling down all the issues raised against its use as an analysis technique [11]. It is a convenient method for analysis of electrical properties of the materials because in this representation electrode polarization effects are suppressed and electrical effects related to bulk can be easily separated [8]. The complex form of electric modulus can be written in frequency domain as [9]:

$$\hat{M}(\omega) = M_\infty \left[ 1 - \int_0^\infty \exp(-i\omega t) \left( \frac{d\Phi}{dt} \right) dt \right] \tag{1}$$

where $\Phi(t)$ describes the relaxation of electric field $E$ after the application of a step in the displacement $D$ and $M_\infty$ is high frequency electrical modulus. The relaxation function is known to follow KWW equation [12,13].

$$\Phi(t) = \phi_0 \exp\left[-\left(\frac{t}{\tau}\right)^{\beta}\right] \qquad (2)$$

where $\beta$ is the stretching parameter and usually takes the value as, $0 < \beta \leq 1$; $\tau$ is the relaxation time and $\phi_0$ is the fraction of the experimental quantity being relaxed. When $\beta = 1$, the relaxation behaviour follows single exponential Debye type mechanism. As the value of $\beta$ becomes lower than 1, the relaxation function $\Phi(t)$ shows more stretching.

The values of $\beta$ may be ascribed to the extent of the interaction between mobile ions present in a glass sample [14]. Decreasing value of $\beta$ is an estimate of increased interaction between the mobile ion species. In mixed alkali glasses where usually an alkali element is gradually replaced by another alkali element, a mismatch in the ionic radii may have its effect on density and hence on the molar volume of the glass. This may open up the glass network. In such cases the interaction between the mobile species should be estimated by not just considering the effect of number of mobile species. A combined effect of the distance between mobile ions and the molar volume may have a greater impact in such cases.

Swenson and Börjesson has shown in past, that the glass network expansion, which is related to the available free volume, is a key parameter determining the increase of the high ionic conductivity in case of fast ionic conductors and cubic scaling relation between the conductivity enhancement and the expansion of the network forming units [15]. Recently, this finding is repeated with Rubidium and Caesium silicate glasses [16] and potassium silicate glasses [17]. Sanson et al, has found out general correlation between the I-Ag distance in AgI fast ion conducting glasses, measured by EXAFS and the glass activation energy for dc ionic conductivity. Glasses with longer I-Ag distances showed higher ionic conductivity, independent of the chemical composition of their host glassy matrix [18]. But the author is unaware of any such attempt to correlate the interaction between mobile ions and the molar volume or distance between mobile ions using electrical modulus formalism.

This communication to the best of authors' knowledge shows for the first time, a corroborative effect of distance between mobile species and molar volume of the glass on the extent of interaction between the mobile ions.

## 2. Experimental

Glasses of general formula $40V_2O_5 - 30BaF_2 - (30-x) LiF - xRbF$ where x = 5, 10,15,20,25 and 30 are prepared via melt quenching technique. The details of the glass preparation and basic characterization is discussed in detail in a previous communication [19]. For a ready reference, Table I summarizes the glass composition along with other parameters like $T_g$ and density of the glass samples.

Large chunks of these samples were polished to get thickness less than 0.5mm. A 5mm diameter mask was used to paint high purity conducting silver (SPI supplies, USA) contact. Real and Imaginary part of the complex impedance was measured using Agilent Precision Impedance Analyzer 4194A over the frequency range of 100 Hz to 10 MHz with an oscillator level of 500 mV which was found to be well within the linear response limit. Temperature dependent measurements were performed using a home built furnace with temperature controller in the range 140 $^o$C to 300 $^o$C; the temperature was measured using a K type thermocouple to an accuracy of $\pm 1$ $^o$C. Complex electrical modulus, $\hat{M} = i2\pi f \varepsilon_0 \hat{Z}(f)/k$, where $\hat{Z}(f)$ - complex impedance, $k$ – cell constant given by $d/A$ ( $d$ is the thickness and $A$ is the area of cross section of the sample) [20] was calculated for each sample and for all the temperature range. Here we report the data from 220 $^o$C and above as the data below 200 $^o$C were noisy

## 3. Results and Discussions

### 3.1. Results

Since it is difficult to extract $\beta$ using above mentioned equations directly due to complex numerical analysis involved, we have used a new function (Eq. (3)) proposed by Bergman [21] to fit the modulus spectra. The proposed equation was originally meant to fit the imaginary part of the general susceptibility $X^*$ to KWW equation but can equally be well approximated to fit imaginary part of the modulus spectra $X''$ for $\beta \geq 0.4$.

$$X''(\omega) = \frac{X''_p}{1-b+\frac{b}{1+b}[b(\omega_p/\omega)+(\omega/\omega_p)^b]} \qquad (3)$$

Where $X''_p$ and $\omega_p$ are peak height and position and $b$ is shape parameter.

The parameters in Eq. (3) are related to KWW parameters as [21]:

$$b \approx \beta$$
$$X''_p \approx (f/2)\beta$$
$$\omega_p \approx 1/[\tau(\beta^{-1}\Gamma(\beta^{-1}))^{1/2}]$$

where $\Gamma$ is the gamma function. Fig. 1 shows the experimental data of frequency dependent imaginary part $M''$ of electrical modulus fitted to Eq. (3) for VBLR3 glass at different temperatures. It can be seen that at high frequencies the fit deviates slightly from the data which may be due to contributions from the constant loss regime [22]. Similar analysis is performed on all other glass samples and the stretching parameter β, extracted from the fitting (at 220 °C) is plotted against Rubidium (Rb) mole fraction in Fig. 2. The vertical bar represents the error involved in the fitting procedure. It can be seen from Fig. 2 that β decreases as Li ions are replaced by Rb and goes through a minimum at 0.5 molar fraction of Rb ions.

The width of the modulus spectra may be quantified by the stretching parameter β. We have observed a decrease in β with respect to rubidium concentration variation and reach a minimum at 50 mol% of rubidium concentration. This is consistent with the general observation that at high alkali ion concentration β decreases [22].

If we consider the time dependence of the dipole moments caused by the ionic vibrations ($10^{-12}$ to $10^{-13}$ s) and electronic motions ($10^{-14}$ to $10^{-15}$ s), the contributions from the two processes is infinitely fast considering the time scale of the impedance experiments (1 μs to 1 s ) [20]. On the other hand, ionic diffusion processes are much slower. The contributions from all these processes contribute to the mean square overall dipole moment. The conductivity can be explicitly divided into conductivity contributions from the fast processes and contributions from the slow processes. As far as conductivity is concerned the fast processes contribute a frequency -independent term, but for the $M''$ spectrum, the shape is influenced by both fast as well as slow processes contributing at high and low frequencies [20].

To estimate the contribution of both fast and slow processes to the total relaxation, we have used Bergman's universal response function [21].

$$X''(\omega) = \frac{X''_p}{\frac{(1-|a-b|)}{a+b}[b(\omega/\omega_p)^{-a} + a(\omega/\omega_p)^{b}] + |a-b|} \qquad (4)$$

Where exponent *a* stands for fast process and *b* stands for the slow process. These exponents take care of the high and low frequency broadening due to fast and slow processes. The extent of coupling between the two processes is measured by |*a* - *b*|. $M''$ spectra of all the samples are fitted with Eq. (4) and coupling between *a* and *b* is estimated. Fig. 3 shows the variation of |*a* - *b*| with mole fraction of Rb atoms. The difference |*a* - *b*| ~ 1, suggest that the contribution from the fast relaxation processes is governing the shape of $M''$ spectrum and contribution from slow process is negligible. From fig. 3, it is seen that when the Li ion content is high and decreasing up to 50 mol%, the contribution from fast process is dominant. As the Rb ions concentration increases, the slow process dominates the relaxation.

In Fig. 4(a) we have shown the scaling of the electrical modulus spectra of VBL glass at different temperatures up to a temperature well above $T_g$. We see that stretching parameter *β* is slightly dependent on temperature. Fig. 4(b) shows the same scaling for all the glass samples, but at a given temperature which is well below the $T_g$. From this plot we see that the modulus spectrum is broader for VBLR3 glass than any other glass. This is consistent with the observation that *β* has a small value for the said glass in comparison with other glasses (refer Fig. 2).

**3.2 Discussions**

**3.2.1 Effect of molar volume on the interaction between mobile ions**

It is well established fact that the stretching parameter *β* increases with decreasing interaction between mobile ions as the concentration of the mobile ions are decreased,[23]. Lower value of *β* signifies strong interaction between the mobile ions. Thus the low value of *β* in Fig. 2 can be attributed to a strong interaction between the mobile ion species.

When studying Mixed Alkali Effect (also known as mixed mobile ion effect) one usually replaces one type of alkali with another, either smaller in size or larger in size than the former alkali. This has a drastic effect on the physical properties of the glasses which involve ion dynamics. In our case, we have replaced smaller alkali (lithium) by a larger alkali (rubidium). This decreases the glass density (see Table I). This is attributed to opening up of the glass network because of the insertion of the larger ion in place of a smaller one. Because of this the molar volume increases. The distance between Rb ions, estimated from knowledge of molar volume and molar mass for a given glass sample, decrease as the concentration of Rb ions increases. These aspects should have contributed to increase the interaction between the mobile ions. But on the contrary, $\beta$ increases at higher concentration of Rb suggesting a reduced interaction (see Fig. 2).

The ratio of distance between Rb ions ($r(Rb\text{-}Rb)$) to the molar volume ($V_m$), i.e., $\frac{r(Rb-Rb)}{V_m}$ is a measure of the **distribution density** of the Rb species for a given glass. In Fig. 5 this ratio is plotted against the fraction of Rb concentration. It is observed that the ratio decreases as the Rb concentration increases.

This observation suggests an explanation to the observed increase in $\beta$ value at higher concentrations of Rb. Single alkali Rb glass (VBR) has the highest molar volume and lowest ratio of $\frac{r(Rb-Rb)}{V_m}$. Also this composition shows the highest value of $\beta$ (see Fig. 2.) indicating the lowest interaction between the mobile ions. Thus it may be concluded that even though the concentration of Rb ions is high and *r(Rb-Rb)* is small, the interaction between Rb ions decrease because the molar volume increases as Rb ion concentration increases. It follows from the above arguments that molar volume has a larger contribution than *r(Rb-Rb)* in deciding the interaction between the mobile ions. It also suggests that at high concentrations of Rb, the mobile ions essentially undergo relaxation independent of each other.

### 3.2.2 Macro and microscopic relaxation of mobile ions

In the light of these arguments, we observe that the decrease in the stretching parameter $\beta$ from its single alkali value is an effect of increased coupling between mobile ions and a large contribution from fast processes denoted by '$a$' in Eq. (4) (Refere Fig. 3). It is found that $\beta$ attains minimum at 0.50 molar fraction of Rb concentration which suggests that there is a good coupling between different relaxing ions/sites and hence all these contribute to macroscopic relaxation given by $\tau_p = \tau_0^* \exp(E_A / k_B T)$ that is responsible for the stretching [1, 24]. Here $E_A$ is the activation energy for thermally activated relaxation. As the concentration of the Rb ions increase, the molar volume increases and the factor $\frac{r(Rb-Rb)}{V_m}$ decreases. This effectively decreases the coupling between the mobile ions and the slow process may start contributing significantly. With increase in $\beta$, decoupling between the relaxing ions/sites increases and they relax with a single microscopic relaxation time, $\tau_0 = \tau_0 \exp(E_A / k_B T)$ and the $M''$ spectra becomes more and more Debye like.

### 4. Conclusions

An electrical modulus analysis of mixed alkali oxyfluoro vanadate glasses is made using Bergman's formula. The variation of the stretching parameter $\beta$, with respect to Rb mole fraction is analysed in the light of variation of distribution density parameter, $\frac{r(Rb-Rb)}{V_m}$ of Rb ions which is the ratio of distance between Rb ions and molar volume. It is observed that the molar volume has a significant contribution in estimating coupling between mobile ions in the glass system studied. We also observed the contribution of distribution density parameter on the macro and microscopic relaxation of the mobile ions.

**5. Acknowledgements:**

The author thanks Prof. K P Ramesh of Dept. of Physics, Indian Institute of Science, Bangalore, India for the help extended in carrying out the experiments.

**Tables and Figures:**

**Table I : Glass compositions, Glass transition temperature (Tg) and Glass density.**

| Sl. No. | Batch Code | Rb/{Li+Rb} in Mol fraction | Glass content (mol %) | | | | $T_g$ (±0.05°C) | Density (±0.01g/cm³) | Molar Volume (±0.01 mol/cm³) | Number of Li ions in $10^{18}$ ions / cc | Average distance between Li ions in $10^{-7}$ cm | Number of Rb ions in $10^{18}$ ions / cc | Average distance between Rb ions in $10^{-7}$ cm |
| --- | --- | --- | --- | --- | --- | --- | --- | --- | --- | --- | --- | --- | --- |
| | | | $V_2O_5$ | $BaF_2$ | LiF | RbF | | | | | | | |
| 1 | VBL | 0 | 40 | 30 | 30 | 0 | 267.25 | 3.68 | 3617.04 | 49.95 | 2.72 | 0 | 0 |
| 2 | VBLR1 | 0.17 | 40 | 30 | 25 | 5 | 263.78 | 3.64 | 3760.52 | 40.04 | 2.92 | 8.01 | 4.99 |
| 3 | VBLR2 | 0.33 | 40 | 30 | 20 | 10 | 264.39 | 3.61 | 3902.55 | 30.87 | 3.19 | 15.43 | 4.02 |
| 4 | VBLR3 | 0.5 | 40 | 30 | 15 | 15 | 262.41 | 3.68 | 3941.42 | 22.92 | 3.52 | 22.92 | 3.52 |
| 5 | VBLR4 | 0.67 | 40 | 30 | 10 | 20 | 261.16 | 3.57 | 4170.77 | 14.41 | 4.11 | 28.88 | 3.26 |
| 6 | VBLR5 | 0.83 | 40 | 30 | 5 | 25 | 260.61 | 3.66 | 4169.79 | 7.22 | 5.17 | 36.11 | 3.03 |
| 7 | VBR | 1 | 40 | 30 | 0 | 30 | 254.29 | 3.45 | 4538.41 | 0 | 0 | 39.81 | 2.93 |

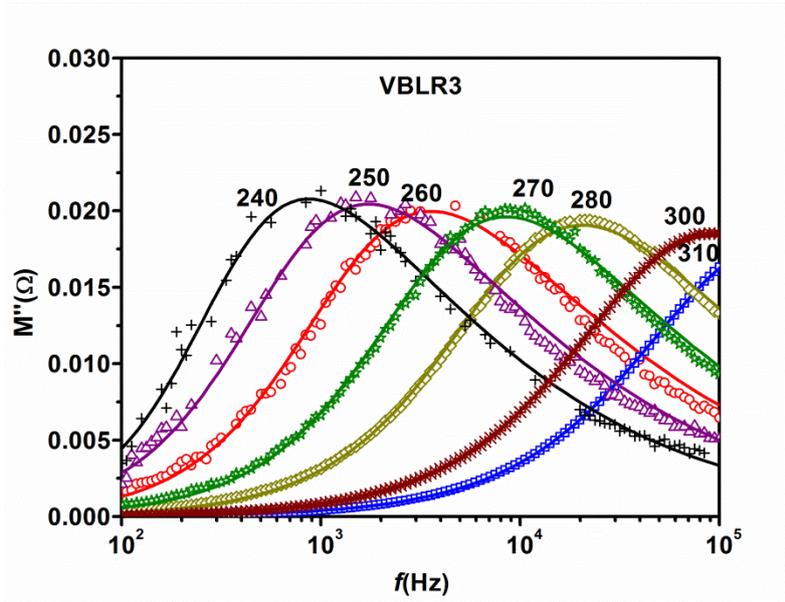

Fig. 1: Electric modulus representation of VBLR3 glass sample. Solid lines are fit to Eq. (3). At high frequency side departure from KWW relation is observed.

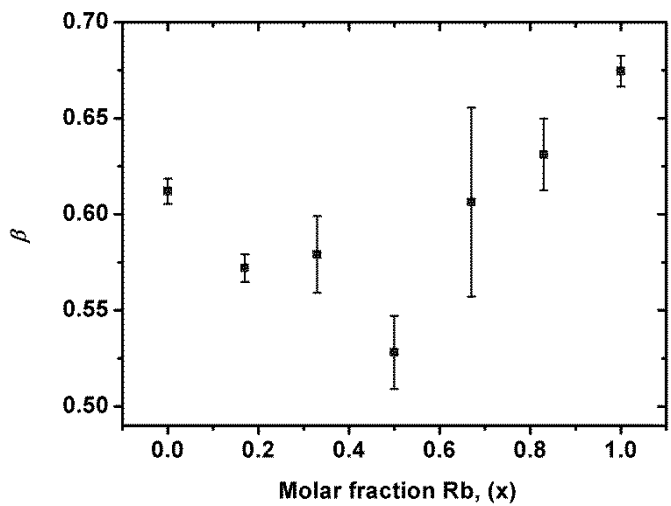

Fig. 2: The KWW stretching parameter $\beta$ vs. Rb molar fraction showing a nonlinear variation.

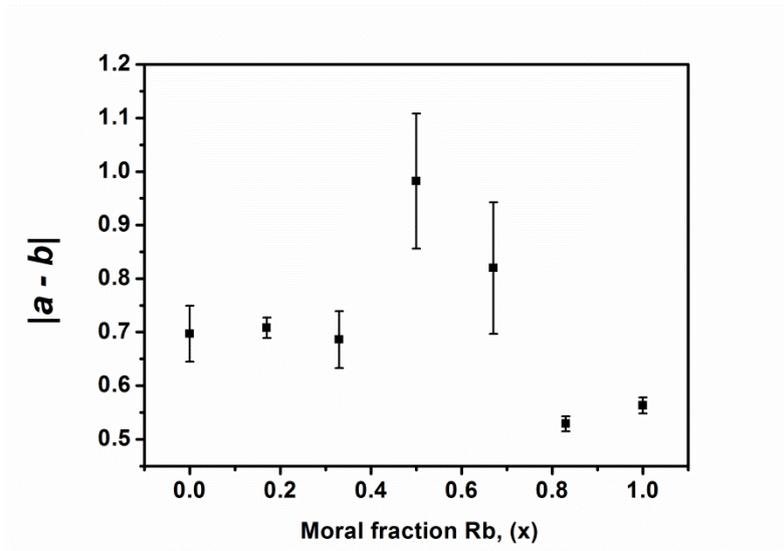

Fig. 3: Variation of coupling between fast and slow processes, |a - b| versus Rb molar fraction. |a - b| is estimated by the fit of M" spectra to Eq.(4). The error bars indicate the uncertainty in |a - b| value.

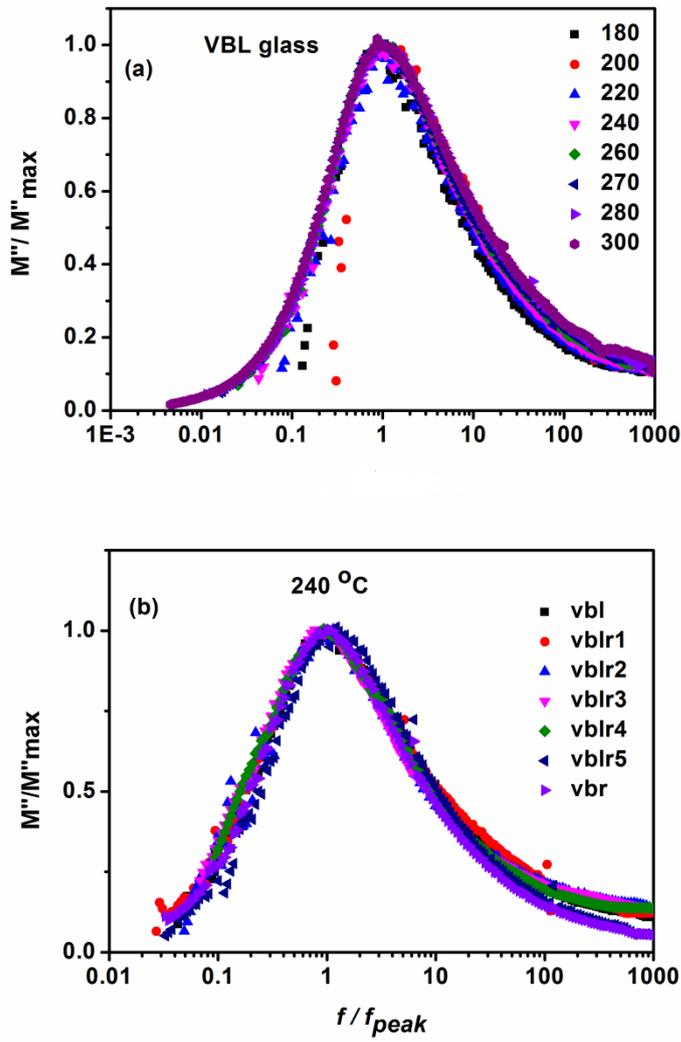

Fig.4: (a) Electrical modulus scaling of VBL glass at different temperatures. $\beta$ shows a slight dependence on the temperature. (b) Electrical modulus scaling of all glass samples at a particular temperature. This temperature is chosen as it is well below the glass transition temperature of all the samples.

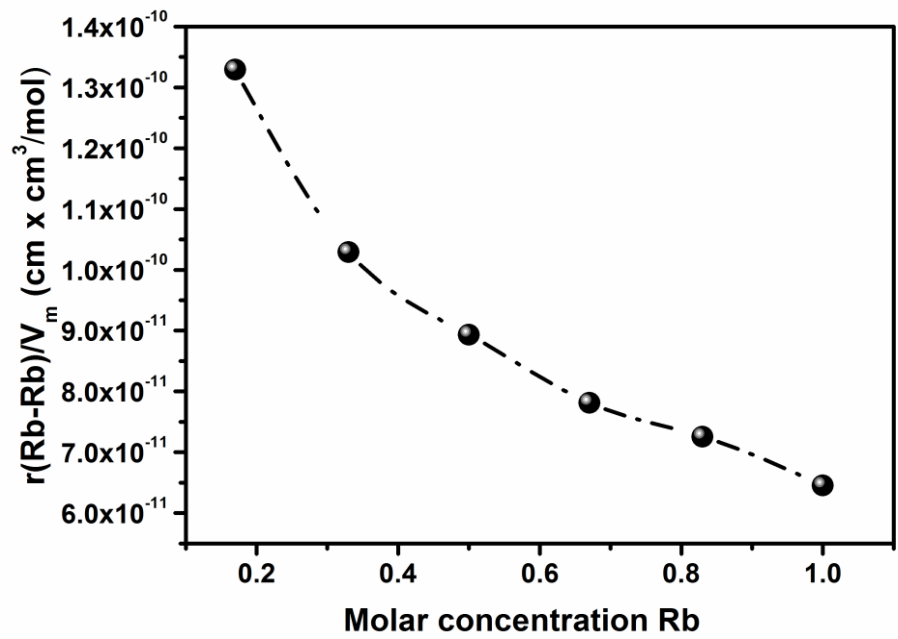

Fig. 5: The change in the distribution density parameter, $\frac{r(Rb-Rb)}{V_m}$ of Rb ions Vs Molar concentration of Rb. The line is drawn as guide to eye.